\documentclass[letterpaper,USenglish,autoref]{lipics-v2021}

\usepackage{amssymb}
\usepackage{stmaryrd}
\usepackage{proof}
\title{Substructural Parametricity}

\author{C. B. Aberl{\'e}}{Carnegie Mellon University}{caberle@andrew.cmu.edu}{}{}
\author{Chris Martens}{Northeastern University}{c.martens@northeastern.edu}{}{}
\author{Frank Pfenning}{Carnegie Mellon University}{fp@cs.cmu.edu}{}{}

\authorrunning{C. B. Aberl{\'e} and Chris Martens and Frank Pfenning}

\Copyright{C. B. Aberl{\'e}, Chris Martens, and Frank Pfenning}



\category{}
\relatedversion{}
\acknowledgements{We would like to thank Karl Crary for an early instantiation of the
  underlying idea.}
\renewcommand{\doiHeading}[3]{}

\newcommand{\ms}[1]{\mathsf{#1}}
\newcommand{\mi}[1]{\mathit{#1}}
\newcommand{\mb}[1]{\mathbf{#1}}

\newcommand{\ums}[1]{\underline{\ms{#1}}}

\newcommand{\arrow}{\mathbin{\rightarrow}}

\newcommand{\tensor}{\otimes}
\newcommand{\lolli}{\multimap}
\newcommand{\fuse}{\bullet}
\newcommand{\lover}{\mathbin{/}}
\newcommand{\lunder}{\mathbin{\backslash}}
\newcommand{\twist}{\mathbin{\circ}}

\newcommand{\one}{\mb{1}}

\newcommand{\semi}{\mathrel{;}}

\newcommand{\mmode}[1]{{\mathchoice{\ms{#1}}{\ms{#1}}{\scriptscriptstyle\ms{#1}}{\scriptscriptstyle\ms{#1}}}}
\newcommand{\mL}{\mmode{L}}

\newcommand{\evalsto}{\mathrel{\hookrightarrow}}
\newcommand{\llb}{\llbracket}
\newcommand{\rrb}{\rrbracket}

\newcommand{\rarrow}{\twoheadrightarrow}
\renewcommand{\lunder}{\rightarrowtail}

\newenvironment{rules}{\[\begin{array}{c}}{\end{array}\]}

\begin{document}

\nolinenumbers

\maketitle

\begin{abstract}
  Ordered, linear, and other substructural type systems allow us to expose deep
  properties of programs at the syntactic level of types.  In this paper, we
  develop a family of unary logical relations that allow us to prove
  consequences of parametricity for a range of substructural type systems.  A
  key idea is to parameterize the relation by an algebra, which we exemplify with a
  monoid and commutative monoid to interpret ordered and linear type systems,
  respectively.  We prove the fundamental theorem of logical relations and apply
  it to deduce extensional properties of inhabitants of certain types.  Examples
  include demonstrating that the ordered types for list append and reversal are
  inhabited by exactly one function, as are types of some tree traversals.
  Similarly, the linear type of the identity function on lists is inhabited only
  by permutations of the input.  Our most advanced example shows that the
  ordered type of the list fold function is inhabited only by the fold function.


\end{abstract}

\section{Introduction}

Substructural type systems and parametric polymorphism are two 
mechanisms for capturing precise behavioral properties of
programs at the type level, enabling powerful static reasoning.
The goal of this paper is to give a theoretical account of these
mechanisms in combination.

Substructural type systems have been investigated since the advent of linear
logic, starting with the seminal paper by Girard and
Lafont~\cite{Girard87tapsoft}.  Among other applications, with substructural
type systems one can avoid garbage collection, update memory in
place~\cite{Lorenzen23icfp,Lorenzen24pldi}, make
message-passing~\cite{Gay10jfp,Caires10concur} or shared memory
concurrency~\cite{Georges25popl,Pfenning23coordination} safe, model quantum
computation~\cite{Fu20lics}, or reason efficiently about imperative
programs~\cite{Lattuada23oopsla}. Substructural type systems have thus been
incorporated into languages that seek to offer such guarantees, such as
Rust, Koka, Haskell, Oxidized OCaml, and ProtoQuipper.

Parametricity, originally introduced for System F~\cite{Reynolds83ip}, enables
the idea that programs whose types involve universal quantification over type
parameters have certain strong semantic properties. This idea supports powerful
program reasoning principles such as representation independence across
abstraction boundaries~\cite{mitchell1986representation} and ``theorems for
free'' that can be derived about all inhabitants of certain types, for example
that every inhabitant of $\forall{\alpha}.\;\alpha \to \alpha$ is equivalent to
the identity function~\cite{Wadler89fpca}.

The theory of substructural logics and type systems is now relatively well
understood, including several ways to integrate substructural and structural
type systems~\cite{Benton94csl,Pruiksma18aun,Jang24fscd}.  It is therefore
somewhat surprising that we do not yet know much about how parametricity and its
applications interact with them.  The main foray into substructural
parametricity is a paper by Zhao et al.~\cite{Zhao10aplas} that accounts for a
polymorphic dual-intuitionistic linear logic.  They point out that logical
relations on closed terms are problematic because substitution obscures
linearity.  Their solution was to construct a logical relation on open terms,
necessitating the introduction of ``semantic typing'' judgments that mirror the
syntactic type system, which complicates their definition and application.

In this paper, we follow an approach using \emph{constructive resource semantics}
in the style of Reed et al.~\cite{Reed07hylo,Reed09phd,Reed10un}
to construct logical relations on \emph{closed terms}.  We start with an ordered
type system~\cite{Polakow99tlca,Polakow01phd,Kanovich18ijcar}, which may be
considered the least permissive among substructural type systems and therefore
admits a pleasantly minimal definition. However, the construction is generic
with respect to certain properties of the resource algebra, which allows us to
extend it also to linear and unrestricted types.
Consequences of our development include that certain polymorphic types are
only inhabited by the polymorphic append and reverse functions on lists.
Similarly, certain types are only inhabited by functions that swap or
maintain
the order of pairs.  The most advanced application shows that the ordered
type of fold over lists is inhabited only by the fold function.

We conjecture that the three substructural modes we investigate---ordered,
linear, and unrestricted---can also be combined in an adjoint
framework~\cite{Benton94csl,Jang24fscd} but leave this to future work.
Similarly, we simplify our presentation by defining only a {\em unary} logical
relation since it is sufficient to demonstrate proof-of-concept, but nothing
stands in the way of a more general definition (for example, to support
representation independence results).

\section{A Minimalist Fragment}
\label{sec:min}

We start with a small fragment of the Full Lambek
Calculus~\cite{Lambek58,MacCaull98jsl}, extended with parametric
polymorphism~\cite{Strachey67}.  This fragment is sufficient to illustrate the
main ideas behind our constructions.  For the sake of simplicity we choose a
Curry-style formulation of typing, concentrating on properties of untyped terms
rather than intrinsically typed terms.  This allows the same terms to inhabit
ordered, linear, and unrestricted types and thereby focus on semantic rather
than syntactic issues.
\[
  \begin{array}{llcll}
    \mbox{Types} & A, B & ::= & \alpha \mid A \fuse B \mid A \lunder B \mid A \rarrow B \mid \forall \alpha.\, A \\
    \mbox{Expressions} & e & ::= & x \\
                 & & \mid & (e_1, e_2) \mid \mb{match}\; e\; ((x, y) \Rightarrow e') & (A \fuse B) \\
                 & & \mid & \lambda x.\ e \mid e_1\, e_2 & (A \lunder B, A \rarrow B) 
  \end{array}
\]
In this fragment, we have $A \fuse B$ (read ``$A$ fuse $B$'') which, logically,
is a noncommutative conjunction.  We have two forms of implication:
$A \lunder B$ (read: ``$A$ under $B$'', originally written as
$A \mathbin{\backslash} B$) which is true if from the hypothesis $A$
\emph{placed at the left end of the antecedents} we can deduce $B$, and
$A \rarrow B$ (read: ``$B$ over $A$'', originally written as $B \mathbin{/} A$)
which is true if from the hypothesis $A$ \emph{placed at the right and of the
  antecedents} we can prove $B$.  Lambek's original notation was suitable for
the sequent calculus and its applications in linguistics, but is less readable
for natural deduction and functional programming.

Our basic typing judgment has the form $\Delta \mid \Omega \vdash e : A$ where
$\Delta$ consists of hypotheses $\alpha \; \ms{type}$, and $\Omega$ is an
\emph{ordered context} $(x_1 : A_1) \ldots (x_n : A_n)$.  We make the standard
presuppositions that $\Delta \vdash A\; \ms{type}$ and
$\Delta \vdash A_i\; \ms{type}$ for every $x_i : A_i$ in $\Omega$, and that both
type variables and term variables are pairwise distinct.  The rules
are show in \autoref{fig:nd}.

\begin{figure}[ht]
\begin{rules}
  \infer[\ms{hyp}]{\Delta \mid x : A \vdash x : A}{}
  \\[1em]
  \infer[{\rarrow}I]
  {\Delta \mid \Omega \vdash \lambda x.\, e : A \rarrow B}
  {\Delta \mid \Omega\, (x : A) \vdash e : B}
  \hspace{3em}
  \infer[{\rarrow}E]
  {\Delta \mid \Omega\, \Omega_A \vdash e_1\, e_2 : B}
  {\Delta \mid \Omega \vdash e_1 : A \rarrow B
    & \Delta \mid \Omega_A \vdash e_2 : A}
  \\[1em]
  \infer[{\lunder}I]
  {\Delta \mid \Omega \vdash \lambda x.\, e : A \lunder B}
  {\Delta \mid (x:A)\, \Omega \vdash e : B}
  \hspace{2em}
  \infer[{\lunder}E]
  {\Delta \mid \Omega_A\, \Omega \vdash e_1\, e_2 : B}
  {\Delta \mid \Omega \vdash e_1 : A \lunder B
    & \Delta \mid \Omega_A \vdash e_2 : A}
  \\[1em]
  \infer[{\fuse}I]
  {\Delta \mid \Omega_A\, \Omega_B \vdash (e_1,e_2) : A \fuse B}
  {\Delta \mid \Omega_A \vdash e_1 : A
    & \Delta \mid \Omega_B \vdash e_2 : B}
  \hspace{1em}
  \infer[{\fuse}E]
  {\Delta \mid \Omega_L\, \Omega\, \Omega_R \vdash \mb{match}\; e\; ((x, y) \Rightarrow e') : C}
  {\Delta \mid \Omega \vdash e : A \fuse B
    & \Delta \mid \Omega_L\, (x : A)\, (y : B)\, \Omega_R \vdash e' : C}
  \\[1em]
  \infer[{\forall}I]
  {\Delta \mid \Omega \vdash e : \forall \alpha.\, A}
  {\Delta, \alpha\, \ms{type} \mid \Omega \vdash e : A}
  \hspace{3em}
  \infer[{\forall}E]
  {\Delta \mid \Omega \vdash e : A(B)}
  {\Delta \mid \Omega \vdash e : \forall \alpha. A(\alpha)
    & \Delta \vdash B\; \ms{type}}
\end{rules}
  \caption{Ordered Natural Deduction}
  \label{fig:nd}
\end{figure}

Here are a few example judgments that hold or fail.   We elide the
context $\Delta =$ $(\alpha\, \ms{type}, \beta\, \ms{type}, \gamma\,
\ms{type})$.

\[
  \begin{array}{lcll}
    & \vdash & \lambda x.\, x : \alpha \lunder \alpha \\
    & \vdash & \lambda x.\, x : \alpha \rarrow \alpha \\
    & \not \vdash & \lambda x.\, \lambda y.\, x : \alpha \rarrow (\beta \rarrow \alpha) & (\mbox{no weakening}) \\
    & \not \vdash & \lambda x.\, (x, x) : \alpha \rarrow (\alpha \fuse \alpha) & (\mbox{no contraction}) \\
    & \vdash & \lambda x.\, \lambda y.\, (x, y) : \alpha \rarrow (\beta \rarrow (\alpha \fuse \beta)) \\
    & \not \vdash & \lambda x.\, \lambda y.\, (x, y) : \alpha \lunder (\beta \lunder (\alpha \fuse \beta)) & (\mbox{no exchange})
    \\[1ex]
    f : \beta \rarrow (\alpha \lunder \gamma)
    & \vdash & \lambda x.\, \lambda y.\, (f\, y)\, x : \alpha \lunder (\beta \rarrow \gamma)
    & (\mbox{``associativity''}) \\
    g : \alpha \lunder (\beta \rarrow \gamma)
    & \vdash &
    \lambda y.\, \lambda x.\, (g\, x)\, y : \beta \rarrow (\alpha \lunder \gamma)
    \\[1ex]
    g : (\alpha \fuse \beta) \rarrow \gamma
    & \vdash & \lambda x.\, \lambda y.\, g\, (x, y) : \alpha \rarrow (\beta \rarrow \gamma)
    & (\mbox{currying}) 
    \\
    f : \alpha \rarrow (\beta \rarrow \gamma)
    & \vdash & \lambda p.\, \mb{match}\; p\; ((x,y) \Rightarrow f\, x\, y) : 
                (\alpha \fuse \beta) \rarrow \gamma
    & (\mbox{uncurrying}) 
  \end{array}
\]

The strictures of the typing judgment imply that certain types may be
uninhabited, or may be inhabited by terms that are extensionally equivalent to a
small number of possibilities.  To count the number of linear functions,
translate $(A \rarrow B)^\mL = (A \lunder B)^\mL = A^\mL \lolli B^\mL$ and
$(A \fuse B)^\mL = A^\mL \tensor B^\mL$ and similarly for unrestricted
functions.
\begin{center}
  \begin{tabular}{lccc}
    Types & Ordered & Linear & Unrestricted \\
    $\alpha \rarrow \alpha$ & 1 & 1 & 1 \\
    $\alpha \rarrow (\alpha \rarrow \alpha)$ & 0 & 0 & 2 \\
    $\alpha \rarrow (\alpha \rarrow (\alpha \fuse \alpha))$ & 1 & 2 & 4 \\
    $\alpha \rarrow (\alpha \lunder (\alpha \fuse \alpha))$ & 1 & 2 & 4\\
    $\alpha \rarrow (\beta \rarrow (\beta \fuse \alpha))$ & 0 & 1 & 1 \\
    $\alpha \rarrow (\beta \rarrow (\alpha \fuse \beta))$ & 1 & 1 & 1
  \end{tabular}
\end{center}

Because our intended application language based on adjoint natural
deduction~\cite{Jang24fscd} is call-by-value, we can give a straightforward
big-step operational semantics~\cite{Kahn87} relating an expression to its final
value.  Because this evaluation does not directly interact with or
benefit from substructural properties, we show it without further comment in
\autoref{fig:eval}.  It has the property of preservation that if
$\cdot \vdash e : A$ and $e \evalsto v$ then $\cdot \vdash v : A$.  Jang et
al. give an account~\cite{Jang24fscd} that exploits linearity
and other substructural properties, although not the lack of exchange.
\begin{figure}[ht]
  \begin{rules}
    \infer[]
    {\lambda x.\, e \evalsto \lambda x.\ e}
    {}
    \hspace{3em}
    \infer[]
    {e_1\, e_2 \evalsto v}
    {e_1 \evalsto \lambda x.\, e_1'
      & e_2 \evalsto v_2
      & [v_2/x]e_1' \evalsto v}
    \\[1em]
    \infer[]
    {(e_1, e_2) \evalsto (v_1, v_2)}
    {e_1 \evalsto v_1
      & e_2 \evalsto v_2}
    \hspace{3em}
    \infer[]
    {\mb{match}\; e\; ((x, y) \Rightarrow e') \evalsto v'}
    {e \evalsto (v_1, v_2)
      & [v_1/x, v_2/y]e' \evalsto v'}
  \end{rules}
  \caption{Big-Step Operational Semantics}
  \label{fig:eval}
\end{figure}

\section{An Algebraic Logical Predicate}
\label{sec:lr}

Because of our particular setting, we define two mutually dependent logical
predicates: $\llb A\rrb$ for closed expressions and $[A]$ for closed values.  In
addition, the relation is parameterized by elements from an algebraic domain
which may have various properties.  For the ordered case, it should be a monoid,
for the linear case a commutative monoid.  However, the rules themselves do not
require this for the pure sets of terms.  We use $m \cdot n$ for the binary
operation on the monoid, and $\epsilon$ for its unit.

Ignoring polymorphism for now, we write $m \Vdash e \in \llb A\rrb$ and
$m \Vdash v \in [A]$, which is defined by
\[
  \begin{array}{lcl}
    m \Vdash e \in \llb A\rrb & \Longleftrightarrow & e \evalsto v \land m \Vdash v \in [A] \\[1em]
    m \Vdash v \in [1] & \Longleftrightarrow & m = \epsilon \land v = (\,) \\
    m \Vdash v \in [A \fuse B] & \Longleftrightarrow & \exists m_1, m_2.\; m = m_1 \cdot m_2 \land v = (v_1, v_2) 
                                                         \land m_1 \Vdash v_1 \in [A] \land m_2 \Vdash v_2 \in [B] \\
    m \Vdash v \in [A \rarrow B] & \Longleftrightarrow & \forall k.\, k \Vdash w \in [A] \Longrightarrow 
                                                        m \cdot k \Vdash v\, w \in \llb B\rrb \\
    m \Vdash v \in [A \lunder B] & \Longleftrightarrow & \forall k.\, k \Vdash w \in [A] \Longrightarrow
                                                         k \cdot m \Vdash v\, w \in \llb B\rrb
  \end{array}
\]
We can see how the algebraic structure of the monoid tracks information about
order if its operation is not commutative.

The key step, as usual in logical predicates of this nature, is the case for
universal quantification and type variables.  We map type variables $\alpha$ to
relations $R_B$ between monoid elements and values in $[B]$ where $B$ is a
closed type.  We indicate this mapping from type variables to sets of values $S$
and write it as a superscript on $\Vdash$.
\[
  \begin{array}{lcl}
    m \Vdash^S v \in [\alpha] & \Longleftrightarrow & m \mathrel{S(\alpha)} v \\
    m \Vdash^S v \in [\forall \alpha.\, A(\alpha)] & \Longleftrightarrow
    & \forall B, R_B.\, m \Vdash ^{S, \alpha \mapsto R_B} v \in [A(\alpha)]
  \end{array}
\]
The mapping $S$ is just passed through identically in the cases of the relation
defined above.

We can already verify some interesting properties.  As a first example we show
that the logical predicates are nonempty.

\begin{theorem}
  \label{ex:isin}
\[
  \epsilon \Vdash \lambda x.\, \lambda y.\, (x, y) \in \llb \forall \alpha.\, \alpha \rarrow (\alpha \rarrow (\alpha \fuse \alpha)) \rrb
\]
\end{theorem}
\begin{proof}
Because the $\lambda$-expression is a value, we need to check
\[
  \epsilon \Vdash \lambda x.\, \lambda y.\, (x, y) \in [\forall \alpha.\, \alpha \rarrow (\alpha \rarrow (\alpha \fuse \alpha))]
\]
By definition, this is true if for an arbitrary $A$ and relation $m \mathrel{R_A} v$ we have
\[
  \epsilon \Vdash^{\alpha \mapsto R_A} \lambda x.\, \lambda y.\, (x, y) \in [\alpha \rarrow (\alpha \rarrow (\alpha \fuse \alpha))]
\]
Using the definition of the logical predicate for right implication twice and
one intermediate step of evaluation, this holds iff
\[
  m \cdot k \Vdash^{\alpha \mapsto R_A} (\lambda y.\, (v, y))\; w \in \llb \alpha \fuse \alpha\rrb 
\]
for all $m, k$ with $m \Vdash^{\alpha \mapsto R_A} v$ and $k \Vdash^{\alpha \mapsto R_A} w$.
By evaluation, this is true iff
\[
  m \cdot k \Vdash^{\alpha \mapsto R_A} (v, w) \in [\alpha \fuse \alpha]
\]
Now we can apply the definition of $[A \fuse B]$, splitting $m \cdot k$ into $m$
and $k$ and reducing it to
\[
  m \Vdash^{\alpha \mapsto R_A} v \land k \Vdash^{\alpha \mapsto R_A} w
\]
Both of these hold because, by assumption, $m \mathrel{R_A} v$ and $k \mathrel{R_A} w$.
\end{proof}

More interesting, perhaps, is the reverse.

\begin{theorem}
  \label{ex:mustbe}
If
\[
  \epsilon \Vdash e \in \llb \forall \alpha.\, \alpha \rarrow (\alpha \rarrow (\alpha \fuse \alpha))\rrb
\]
then $e$ is extensionally equal to $\lambda x.\, \lambda y.\, (x, y)$.
In particular, it can not be $\lambda x.\, \lambda y.\, (y, x)$.
\end{theorem}
\begin{proof}
We choose our monoid to be the free monoid over two generators $a$ and $b$ and we
choose an arbitrary closed type $A$ and two elements $v$ and $w$.  Moreoever, we
pick $R_A$ relating only $a \mathrel{R_A} v$ and $b \mathrel{R_A} w$.

From the definitions (and skipping over some simple properties regarding
evaluation), we obtain
\[
  a \cdot b \Vdash^{\alpha \mapsto R_A} e\; v\; w \in \llb \alpha \fuse \alpha\rrb
\]
By the clauses for $\llb \alpha \fuse \alpha\rrb$, $[\alpha \fuse \alpha]$ and
$\alpha$ we conclude that
\[
  e\; v\; w \evalsto (u_1, u_2) 
\]
for some values $u_1$ and $u_2$ with $a \mathrel{R_A} u_1$ and
$b \mathrel{R_A} u_2$.  Because the only value related to $a$ is $v$ and
the only value related to $b$ is $w$, we conclude $u_1 = v$ and $u_2 = w$.
Therefore
\[
  e\; v\; w \evalsto (v, w)
\]
Since $v$ and $w$ were chosen arbitrarily, we see that $e$ is extensionally
equal to $\lambda x.\, \lambda y.\, (x, y)$.
\end{proof}

\section{The Fundamental Theorem}
\label{sec:ftlr}

The fundamental theorem of logical predicates states that every well-typed term
is in the predicate.  Our relations also include terms that are not well-typed,
which can occasionally be useful when one exceeds the limits of static typing.

We need a few standard lemmas, adapted to this case.  We only spell out
one.

\begin{lemma}[Compositionality]
  \label{lm:compositionality}
  Define $R_A$ such that $k \mathrel{R_A} w$ iff $k \Vdash w \in [A]$.  Then
  $m \Vdash^{S, \alpha \mapsto R_A} v \in [B(\alpha)]$ iff $m \Vdash^S v \in [B(A)]$
\end{lemma}
\begin{proof}
  By induction on $B(\alpha)$.
\end{proof}

We would like to prove the fundamental theorem by induction over the structure
of the typing derivation.  Since our logical relation is defined for closed
terms, we need a closing substitution $\eta$.  We define:
\[
  \begin{array}{lcl}
    m \Vdash^S (x \mapsto v) \in [x : A]
    & \Longleftrightarrow
    & m \Vdash^S v \in [A] \\
    m \Vdash^S (\eta_1\, \eta_2) \in [\Omega_1\, \Omega_2]
    & \Longleftrightarrow
    & \exists m_1, m_2.\, m = m_1\cdot m_2 \land m_1 \Vdash^S \eta_1 \in [\Omega_1]
      \land m_1 \Vdash^S \eta_2 \in [\Omega_2]
    \\
    m \Vdash^S (\cdot) \in [\cdot]
    & \Longleftrightarrow
    & m = \epsilon
  \end{array}
\]
Due to the associativity of the monoid operation and concatenation of contexts,
this constitutes a valid definition.

\begin{theorem}[Fundamental Theorem (purely ordered)]
  Assume $\Delta \mid \Omega \vdash e : A$, a mapping $S$ with domain $\Delta$,
  and closing substitution $m \Vdash^S \eta \in [\Omega]$.
  Then $m \Vdash^S \eta(e) \in \llb A\rrb$.
\end{theorem}
\begin{proof}
  By induction on the structure of the given typing derivation.  We show
  a few cases.
  \begin{description}
  \item[Case:]
    \[
      \infer[\ms{hyp}]{\Delta \mid x : A \vdash x : A}{}
    \]
    Then $m \Vdash^S \eta(x) \in [A]$ by assumption and definition, and
    $m \Vdash^S \eta(x) \in \llb A\rrb$ since $\eta(x)$ is a value.
  \item[Case:]
    \[
      \infer[{\rarrow}I]
      {\Delta \mid \Omega \vdash \lambda x.\, e : A \rarrow B}
      {\Delta \mid \Omega\, (x : A) \vdash e : B}
    \]
    \begin{tabbing}
      $m \Vdash^S \eta \in [\Omega]$ \` Given \\
      $k \Vdash^S v \in [A]$ \` Assumption (1) \\
      $k \Vdash^S (x \mapsto v) \in [x : A]$ \` By definition \\
      $m \cdot k \Vdash^S (\eta, x \mapsto v) \in [\Omega\, (x : A)]$ \` By definition \\
      $m \cdot k \Vdash^S (\eta, x \mapsto v)(e) \in \llb B\rrb$ \` By ind.\ hyp. \\
      $m \cdot k \Vdash^S (\eta (\lambda x.\, e))\, v \in \llb B\rrb$ \` By reverse evaluation, $v$ closed \\
      $m \Vdash^S \eta(\lambda x.\, e) \in [A \rarrow B]$ \` By definition, discharging (1) \\
      $m \Vdash^S \eta(\lambda x.\, e) \in \llb A \rarrow B\rrb$ \` By definition
    \end{tabbing}
  \item[Case:]
    \[
      \infer[{\rarrow}E]
      {\Delta \mid \Omega\, \Omega_A \vdash e_1\, e_2 : B}
      {\Delta \mid \Omega \vdash e_1 : A \rarrow B
        & \Delta \mid \Omega_A \vdash e_2 : A}
    \]
    \begin{tabbing}
      $m \Vdash^S \eta \in [\Omega\, \Omega_A]$ \` Given \\
      $m_1 \Vdash^S \eta_1 \in [\Omega]$ and $m_2 \Vdash^S \eta_2 \in [\Omega_A]$ \\
      for some $m_1$, $m_2$, $\eta_1$, and $\eta_2$
      with $m = m_1 \cdot m_2$ and $\eta = \eta_1\; \eta_2$ \` By definition \\
      $m_1 \Vdash^S \eta_1(e_1) \in \llb A \rarrow B\rrb$ \` By ind.\ hyp. \\
      $m_2 \Vdash^S \eta_2(e_2) \in \llb A\rrb$ \` By ind.\ hyp. \\
      $\eta_1(e_1) \evalsto v_1$ with $m_1 \Vdash^S v_1 \in [A \rarrow B]$ \` By definition \\
      $\eta_2(e_2) \evalsto v_2$ with $m_2 \Vdash^S v_2 \in [A]$ \` By definition \\
      $m_1\cdot m_2 \Vdash^S v_1\, v_2 \in \llb B\rrb$ \` By definition \\
      $(\eta_1\, \eta_2)(e_1\, e_2) = (\eta_1(e_1))\, (\eta_2(e_2))$ \` By properties of substitution \\
      $m \Vdash^S \eta(e_1\, e_2) \in \llb B\rrb$ \` Since $m = m_1\cdot m_2$ and $\eta = (\eta_1\, \eta_2)$
    \end{tabbing}
  \item[Case:]
    \[
      \infer[{\forall}I]
      {\Delta \mid \Omega \vdash e : \forall \alpha.\, A}
      {\Delta, \alpha\, \ms{type} \mid \Omega \vdash e : A}
    \]
    \begin{tabbing}
      $m \Vdash^S \eta \in [\Omega]$ \` Given \\
      $R_B$ an arbitrary relation $k \mathrel{R_B} v$ \` Assumption (1) \\
      $m \Vdash^{S, \alpha \mapsto R_B} \eta \in [\Omega]$ \` Since $\alpha$ fresh \\
      $m \Vdash^{S, \alpha \mapsto R_B} \eta(e) \in \llb A\rrb$ \` By ind.\ hyp. \\
      $m \Vdash^S \eta(e) \in \llb \forall \alpha.\, A\rrb$ \` By definition, discharging (1)
    \end{tabbing}
  \item[Case:]
    \[
      \infer[{\forall}E]
      {\Delta \mid \Omega \vdash e : A(B)}
      {\Delta \mid \Omega \vdash e : \forall \alpha. A(\alpha)
        & \Delta \vdash B\; \ms{type}}
    \]
    \begin{tabbing}
      $m \Vdash^S \eta \in [\Omega]$ \` Given \\
      $m \Vdash^S \eta(e) \in \llb \forall \alpha.\, A(\alpha)\rrb$ \` By ind.\ hyp. \\
      Define $k \mathrel{R}_B v$ iff $k \Vdash^S v \in [B]$ \\
      $m \Vdash^{S, \alpha \mapsto R_B} \eta(e) \in \llb A(\alpha)\rrb$ \` By definition \\
      $m \Vdash^{S, \alpha \mapsto R_B} v \in [A(\alpha)]$ for $\eta(e) \evalsto v$ \` By definition \\
      $m \Vdash^S v \in [A(B)]$ \` By compositionality (Lemma~\ref{lm:compositionality}) \\
      $m \Vdash^S \eta(e) \in \llb A(B)\rrb$ \` By definition 
    \end{tabbing}
  \end{description}
\end{proof}

Because typing implies that the logical predicate holds, the
earlier examples now apply to well-typed terms.
\begin{theorem}[(\autoref{ex:mustbe} revisited)]
  If
  \[
    \cdot \vdash e : \forall \alpha.\, \alpha \rarrow (\alpha \rarrow (\alpha \fuse \alpha))
  \]
  then $e$ is extensionally equivalent to $\lambda x.\, \lambda y.\, (x, y)$.
\end{theorem}
\begin{proof}
  We just note that
  \[
    \epsilon \Vdash e \in \llb \forall \alpha.\, \alpha \rarrow (\alpha \rarrow (\alpha \fuse \alpha))\rrb
  \]
  since $(\cdot) \in [\cdot]$ and $(\cdot)e = e$ and the empty mapping $S$ suffices
  without any free type variables.  Then we appeal to the reasoning in \autoref{ex:mustbe}.
\end{proof}

\section{Unrestricted Functions}
\label{sec:u}

We are interested in properties of functions such as list append or list reversal,
or higher-order functions such as fold.  This requires inductive types, but the
functions on them are not used linearly.  For example, append has a recursive
call in the case of a nonempty list, but none in the case of an empty list.  We
could introduce a general modality ${!}A$ for this purpose.  A simpler
alternative that is sufficient for our situation is to introduce unrestricted
function types $A \arrow B$ (usually coded as ${!}A \lolli B$ in linear logic or
${!}A \rarrow B$ in ordered logic).  This path has been explored
previously~\cite{Polakow99tlca} with different motivations.  There, an
\emph{open} logical relation was defined on the negative monomorphic fragment in
order to show the existence of canonical forms, a property that is largely
independent of ordered typing.

Adding unrestricted functions is rather straightforward in typing by using two
kinds of variables: those that are ordered and those unrestricted.  Then, in the
logical predicate, unrestricted variables must not use any resources, that is,
they are assigned the unit element $\epsilon$ of the monoid during the definition.

The generalized judgment has the form
$\Delta \mid \Gamma \semi \Omega \vdash e : A$ where $\Gamma$ contains type
assignments for variables that can be used in an unrestricted (not linear and
not ordered) way.  All the previous rules are augmented by propagating $\Gamma$
from the conclusion to all premises.  Because our term language is untyped, no
extensions are needed there.  Similarly, the rules of our dynamics do not need
to change.

\begin{figure}[ht]
\begin{rules}
  \infer[\ms{hyp}]{\Delta \mid \Gamma, x : A \semi \cdot \vdash x : A}{}
  \\[1em]
  \infer[{\arrow}I]
  {\Delta \mid \Gamma \semi \Omega \vdash \lambda x.\, e : A \arrow B}
  {\Delta \mid \Gamma, x : A \semi \Omega \vdash e : B}
  \hspace{3em}
  \infer[{\arrow}E]
  {\Delta \mid \Gamma \semi \Omega \vdash e_1\, e_2 : B}
  {\Delta \mid \Gamma \semi \Omega \vdash e_1 : A \arrow B
    & \Delta \mid \Gamma \semi \cdot \vdash e_2 : A}
\end{rules}
\caption{Unrestricted functions}
\label{fig:u}
\end{figure}

We extend the logical predicate using arguments not afforded any resources.
\[
  \begin{array}{lcl}
    m \Vdash v \in [A \arrow B] & \Longleftrightarrow & \forall w.\, \epsilon \Vdash w \in [A] \Longrightarrow
                                                        m \Vdash v\, w \in \llb B\rrb \\
  \end{array}
\]
The fundamental theorem extends in a straightforward way.

\begin{theorem}[Fundamental Theorem (mixed ordered/unrestricted)]
  Assume $\Delta \mid \Gamma \semi \Omega \vdash e : A$, a mapping $S$ with
  domain $\Delta$, and two closing substitutions
  $\epsilon \Vdash^S \theta \in [\Gamma]$ and $m \Vdash^S \eta \in [\Omega]$.  Then
  $m \Vdash^S (\theta \semi \eta)(e) \in \llb A\rrb$.
\end{theorem}
\begin{proof}
  By induction on the structure of the given typing derivation.
\end{proof}

An interesting side effect of these definitions is that if we omit ordered
functions but retain pairs we obtain the ``usual'' formulation closed logical
predicates, including certain consequences of parametricity for the
ordinary $\lambda$-calculus.

\begin{theorem}
  If
  \[
    \cdot \vdash e : \forall \alpha.\, \alpha \arrow (\alpha \arrow (\alpha \fuse \alpha))
  \]
  then $e$ is extensionally equivalent to one of 4 functions: $\lambda x.\, \lambda y.\, (x, y)$,
  $\lambda x.\, \lambda y.\, (y, x)$, $\lambda x.\, \lambda y.\, (x, x)$, or $\lambda x.\, \lambda y.\, (y, y)$.
\end{theorem}
\begin{proof}
  By the fundamental theorem, we have
  \[
    \epsilon \Vdash e \in \llb \forall \alpha.\, \alpha \arrow (\alpha \arrow (\alpha \fuse \alpha)) \rrb
  \]
  We use this for an abitrary closed type $A$ with two arbitary values $v$, and
  $w$ and relation $R_A$ with $\epsilon \mathrel{R_A} v$ and $\epsilon \mathrel{R_A} w$.  Exploiting the
  definition, we get
  \[
    \epsilon \Vdash^{\alpha \mapsto R_A} e \in \llb \alpha \arrow (\alpha \arrow (\alpha \fuse \alpha))\rrb
  \]
  Using the definition of function twice and skipping over some evaluation and
  reverse evaluation, we obtain
  \[
    \epsilon \Vdash^{\alpha \mapsto R_A} f\, v\, w \in \llb \alpha \fuse \alpha\rrb
  \]
  This means that $f\, v\, w \evalsto (u_1, u_2)$ with $\epsilon \mathrel{R_A} u_1$
  and $\epsilon \mathrel{R_A} u_2$.  Because of the definition of $R_A$ there are 4 possibilities
  for $(u_1, u_2)$, namely $(v, w)$, $(w, v)$, $(v, v)$ and $(w, w)$.  This in turn means
  $e$ is extensionally equal to one of the 4 functions shown.
\end{proof}

\section{Unit, Sums, Twist, and Recursive Types}
\label{sec:rec}

At this point, we are at a crossroads.  Because we would like to prove theorems
regarding more complex data structures such as lists, trees, or streams, we
could extend the development with general inductive and coinductive types and
their recursors.  We conjecture that this is possible and leave it to future
work.  The other path is to work with \emph{purely positive types}, including
recursive ones whose values can be directly observed.  In this approach, the
definition of the logical predicate is quite easy to extend.  It becomes a
nested inductive definition: either the type becomes smaller or, once we
encounter a purely positive type and recursion is possible, from then on the
terms become strictly smaller.  In this paper we take the latter approach, which
excludes coinductive types such as streams from consideration, but still yields
many interesting and intuitive consequences.

We take the opportunity to also round out our language with unit, sums, and
twist (the symmetric counterpart of fuse).  We use a signature defining
\emph{equirecursive type names} that may be arbitrarily mutually recursive.
Because such type definitions are otherwise closed, they constitute
metavariables in the sense of contextual modal type
theory~\cite{Nanevski08tocl}.  Each type definition $F[\Delta] = A^+$ must be
\emph{contractive}, that is, its definiens cannot be be another type name.
Moreover, $A^+$ must be \emph{purely positive}, which is interpreted
\emph{inductively}.
\[
  \begin{array}{llcll}
    \mbox{Types} & A & ::= & \ldots \mid A \twist B \mid \oplus\{\ell : A_\ell\}_{\ell \in L} \mid \one \\
    \mbox{Purely Positive Types} & A^+, B^+ & ::= & A^+ \fuse B^+ \mid A^+ \twist B^+
                                                    \mid \one \mid \oplus\{\ell : A_\ell^+\}_{\ell \in L}
                                                    \mid F[\theta] \\
    \mbox{Type Definitions} & \Sigma & ::= & F[\Delta] = A^+ \mid (\cdot) \mid \Sigma_1, \Sigma_2 \\
    \mbox{Type Substitutions} & \theta & ::= & \alpha \mapsto A^+ \mid (\cdot) \mid \theta_1\, \theta_2 \\[1em]
  \end{array}
\]
The language of expressions does not change much because type names are
equirecursive.
\[
  \begin{array}{llcll}
    \mbox{Expression} & e & ::= & \ldots \\
                 & & \mid & k(e) \mid \mb{match}\; e\; \{\ell(x_\ell) \Rightarrow e'\}_{\ell \in L} & (\oplus\{\ell : A_\ell\}) \\
                 & & \mid & (\,) \mid \mb{match}\; e\; ((\,) \Rightarrow e') & (\one)
  \end{array}
\]
We add the type $A \twist B$ (``twist''), symmetric to $A \fuse B$, since encoding it as
$B \fuse A$ requires rewriting terms, flipping the order of pairs.  For
$A \twist B$ it is merely the typechecking that changes.  This allows more types
to be assigned to the same term.  We allow silent unfolding of type definitions,
so there are no explicit rules for $F[\theta]$.

\begin{figure}[ht]
  \begin{rules}
  \infer[{\twist}I]
  {\Delta \mid \Gamma \semi \Omega_B\, \Omega_A \vdash (e_1,e_2) : A \twist B}
  {\Delta \mid \Gamma \semi \Omega_A \vdash e_1 : A
    & \Delta \mid \Gamma \semi \Omega_B \vdash e_2 : B}
  \\[1em]
  \infer[{\twist}E]
  {\Delta \mid \Gamma \semi \Omega_L\, \Omega\, \Omega_R \vdash \mb{match}\; e\; ((x, y) \Rightarrow e') : C}
  {\Delta \mid \Gamma \semi \Omega \vdash e : A \twist B
    & \Delta \mid \Gamma \semi \Omega_L\, (y : B)\, (x : A)\, \Omega_R \vdash e' : C}
  \\[1em]
  \infer[{\one}I]
  {\Delta \mid \Gamma \semi \cdot \vdash (\,) : \one}
  {}
  \hspace{3em}
  \infer[{\one}E]
  {\Delta \mid \Gamma \semi \Omega_L\, \Omega\, \Omega_R \vdash \mb{match}\; e\; ((\,) \Rightarrow e') : C}
  {\Delta \mid \Gamma \semi \Omega \vdash e : A \twist B
    & \Delta \mid \Gamma \semi \Omega_L\, \Omega_R \vdash e' : C}
  \\[1em]
  \infer[{\oplus}I]
  {\Delta \mid \Gamma \semi \Omega \vdash k(e) : {\oplus}\{\ell : A_\ell\}_{\ell \in L}}
  {(k \in L) & \Delta \mid \Gamma \semi \Omega \vdash e : A_k}
  \\[1em]
  \infer[{\oplus}E]
  {\Delta \mid \Gamma \semi \Omega_L\, \Omega\, \Omega_R \vdash \mb{match}\; e\; \{\ell(x_\ell) \Rightarrow e_\ell\}_{\ell \in L} : C}
  {\Delta \mid \Gamma \semi \Omega \vdash e : {\oplus}\{\ell : A_\ell\}_{\ell \in L}
    & (\Delta \mid \Gamma \semi \Omega_L\, (x_\ell : A_\ell)\, \Omega_R \vdash e_\ell : A_\ell)
    & (\forall \ell \in L)}
  \end{rules}
  \caption{Ordered Natural Deduction, Extended}
\end{figure}

\begin{figure}[ht]
  \begin{rules}
    \infer[]
    {(\,) \evalsto (\,)}
    {}
    \hspace{3em}
    \infer[]
    {\mb{match}\; e\; ((\,) \Rightarrow e') \evalsto v'}
    {e \evalsto (\,)
      & e' \evalsto v'}
    \\[1em]
    \infer[]
    {k(e) \evalsto k(v)}
    {e \evalsto v}
    \hspace{3em}
    \infer[]
    {\mb{match}\; e\; \{\ell(x_\ell) \Rightarrow e_\ell\}_{\ell \in L}
      \evalsto v'}
    {e \evalsto k(v)
      & [v/x_k] e_k \evalsto v'}
  \end{rules}
  \caption{Big-Step Operational Semantics, Extended}
  \label{fig:eval-extended}
\end{figure}

The logical predicate is also extended in a straightforward manner.  We assume
the signature $\Sigma$ is fixed and therefore do not carry it explicitly through
the definitions.
\[
  \begin{array}{lcl}
    m \Vdash^S v \in [\one] & \Longleftrightarrow & m = \epsilon \land v = (\,) \\
    m \Vdash^S v \in [A \twist B] & \Longleftrightarrow & \exists m_1, m_2.\; m = m_2 \cdot m_1 \land v = (v_1, v_2) \\
    & & \null \land m_1 \Vdash^S v_1 \in [A] \land m_2 \Vdash^S v_2 \in [B]
    \\
    m \Vdash^S k(v) \in [{\oplus}\{\ell : A_\ell\}_{\ell \in L}]
                            & \Longleftrightarrow & m \Vdash^S v \in [A_k] \land k \in L
    \\
    m \Vdash^S v \in [F[\theta]]
                            & \Longleftrightarrow
                                                  & m \Vdash^S v \in \theta(A^+) \ \mbox{where $F[\Delta] = A^+ \in \Sigma$}
  \end{array}
\]

Because we have equirecursive type definitions, the last clause is usually
applied silently.  The definition of the logical predicate is no longer
straightforwardly inductive on the structure of the type.  However, we see that for purely
positive types (the only ones involved in recursion), the \emph{value} in the
definition becomes strictly smaller in each clause if type definitions are
contractive.  In other words, we now have a nested inductive definition of the
logical predicate, first on the type, and when the type is purely positive, on
the structure of the value.

We can also add recursion to our expression language with the key proviso that
we either restrict ourselves to certain patterns of recursion (for example,
primitive recursion), or termination is guaranteed by other external means (for
example, using an analysis using sized types~\cite{Abel16jfp}).  This assumption
allows us to maintain the structure of the logical predicate, even if it is no
longer a means to prove termination (which we are not interested in for this
paper).

\begin{lemma}[Compositionality (including purely positive equirecursive types)]
  Define $R_A$ such that $k \mathrel{R_A} w$ iff $k \Vdash w \in [A]$.  Then
  $m \Vdash^{S, \alpha \mapsto R_A} v \in [B(\alpha)]$ iff $m \Vdash^S v
  \in [B(A)]$.
\end{lemma}
\begin{proof}
  By nested induction on the definition of the logical predicate for
  $B(\alpha)$, first on the structure of $B$ and second on the structure of the
  value when a purely positive type $F[\theta]$ has been reached.
\end{proof}

\begin{theorem}[Fundamental Theorem (including purely positive recursive types)]
  Assume $\Delta \mid \Gamma \semi \Omega \vdash e : A$, a mapping $S$ with
  domain $\Delta$, and two closing substitutions
  $\epsilon \Vdash^S \theta \in [\Gamma]$ and $m \Vdash^S \eta \in [\Omega]$.  Then
  $m \Vdash^S (\theta \semi \eta)(e) \in \llb A\rrb$.
\end{theorem}
\begin{proof}
  By induction on the structure of the given typing derivation.  When reasoning
  about functions and recursion, we need the assumption of termination.
\end{proof}

\section{Free Theorems for Ordered Lists}

We start with some theorems about ordered lists, not unlike those analyzed by
Wadler~\cite{Wadler89fpca}, but much sharper due to substructural typing.  We
define two versions of ordered lists, one that is ordered left-to-right and one
that is ordered right-to-left.  Both of these use exactly the same
representation; just their typing is different.
\begin{tabbing}
  \quad $\mi{llist}\; \alpha = {\oplus}\{\ums{nil} : \one, \ums{cons} : \alpha \fuse \mi{llist}\; \alpha\}$ \\
  \quad $\mi{rlist}\; \alpha = {\oplus}\{\ums{nil} : \one, \ums{cons} : \alpha \twist \mi{rlist}\; \alpha\}$
\end{tabbing}
The following will be a useful lemma about ordered lists.
\begin{lemma}[Ordered Lists]
  \label{lm:ordlist}
\[
  \begin{array}{lcl}
    m \Vdash^S v \in [\mi{llist}\; \alpha]
    & \Longleftrightarrow
    & m = \epsilon \land v = \ums{nil}\, (\,) \\
    & & \null \lor \exists m_1, m_2.\, m = m_1\cdot m_2 \land v = \ums{cons}\; (v_1, v_2) \\
    & & \quad \null \land m_1 \mathrel{S(\alpha)} v_1 \land m_2 \Vdash v_2 \in [\mi{llist}\; \alpha]
    \\[1em]
    m \Vdash^S v \in [\mi{rlist}\; \alpha]
    & \Longleftrightarrow
    & m = \epsilon \land v = \ums{nil}\, (\,) \\
    & & \null \lor \exists m_1, m_2.\, m = m_2\cdot m_1 \land v = \ums{cons}\; (v_1, v_2) \\
    & & \quad \null \land m_1 \mathrel{S(\alpha)} v_1 \land m_2 \Vdash v_2 \in [\mi{rlist}\; \alpha]
  \end{array}
\]
\end{lemma}
\begin{proof}
  By unrolling the definitions of the logical predicate and the equirecursive
  nature of the definition of lists.
\end{proof}

For the applications, we abbreviate lists, writing $[v_1, \ldots, v_n]$ for
$\ums{cons}(v_1, \ldots, \ums{cons}(v_n, \ums{nil}\, (\,)))$.
\[
  \begin{array}{l}
    m \Vdash^{\alpha \mapsto R_A} v \in [\mi{llist}\; \alpha] 
    \Longleftrightarrow
    m = m_1 \cdots m_n, v = [v_1, \ldots, v_n]
    \ \mbox{where}\  m_i \mathrel{R_A} v_i\ \mbox{(for some $m_i, v_i$)}
    \\[1em]
    m \Vdash^{\alpha \mapsto R_A} v \in [\mi{rlist}\; \alpha]
    \Longleftrightarrow
    m = m_n \cdots m_1, v = [v_1, \ldots, v_n]
    \ \mbox{where}\  m_i \mathrel{R_A} v_i\ \mbox{(for some $m_i, v_i$)}
  \end{array}
\]

Now we state a first property of lists that follows as a consequence of our parameterized
logical predicate.

\begin{theorem}
  \label{ex:listid}
  If $\cdot \vdash f : \forall \alpha.\, \mi{llist}\; \alpha \rarrow \mi{llist}\; \alpha$
  then $f$ is extensionally equal to the identity function on lists.
\end{theorem}
\begin{proof}
  By the fundamental theorem, we have
  \[
    \epsilon \Vdash f \in [\forall \alpha.\, \mi{llist}\; \alpha \rarrow \mi{llist}\; \alpha]
  \]
  To construct a relation $R_A$ we pick an arbitary closed type $A$.  For the
  monoid, we pick the one freely generated by $a_1, a_2, \ldots$ and define
  \[
    m \mathrel{R_A} v \Longleftrightarrow m = a_i \land v = v_i
  \]
  for arbitrary elements $v_i$.   By definition, we obtain
  \[
    \epsilon \Vdash^S f \in [\mi{llist}\; \alpha \rarrow \mi{llist}\; \alpha]
  \]
  Again by definition, that's the case iff
  \[
    \forall m, v.\, m \Vdash^S v \in [\mi{llist}\; \alpha] \Longrightarrow \epsilon \cdot m \Vdash^S f\, v \in \llb \mi{llist}\; \alpha\rrb 
  \]
  Here, $\epsilon \cdot m = m$, by the monoid laws.  Therefore $f\, v \evalsto w$ and
  \[
    \forall m, v.\, m \Vdash^S v \in [\mi{llist}\; \alpha] \Longrightarrow m \Vdash^S w \in [\mi{llist}\; \alpha]
  \]
  We use this for $m = a_1 \cdots a_n$ and $v = [v_1, \ldots, v_n]$.  By our
  lemma about lists and the arbitrary nature of $A$ and $v_i$ we conclude that $w = v$.
\end{proof}

By similar reasoning we can obtain the following properties.
\begin{theorem}[]\mbox{}
  \begin{enumerate}
  \item If $f : \forall \alpha. \mi{rlist}\; \alpha \rarrow \mi{rlist}\; \alpha$ then
    $f$ is extensionally equal to the identity function.
  \item If $f : \forall \alpha.\mi{rlist}\; \alpha \rarrow \mi{llist}\; \alpha$ then
    $f$ is extensionally equal to the list reversal function.
  \item If $f : \forall \alpha.\mi{llist}\; \alpha \rarrow \mi{rlist}\; \alpha$
    then $f$ is extensionally equal to the list reversal function.
  \end{enumerate}
\end{theorem}
\begin{proof}
  By very similar reasoning to the one in \autoref{ex:listid}.
\end{proof}

But can we deduce properties of higher-order functions using ordered
parametricity?  We show one primary example; others such as $\mi{map}$ follow
directly from it or similarly.

Unlike the usual or even linear parametricity, the type of $\mi{fold}$
guarantees that it must be \emph{the} fold function!  Note that the combining
function and initial element are unrestricted arguments (one is called for every
list element, and one is called only for the empty list), but that the combining function's
arguments are ordered.

\begin{theorem}
  If
  \[
    \cdot \vdash f : \forall \alpha.\, \forall \beta.\, (\alpha \fuse \beta \rarrow \beta)
    \arrow \beta \arrow \mi{llist}\; \alpha \rarrow \beta
  \]
  then $f$ extensionally equal to the fold function, that is,
  \[
    f\; g\; b\; [v_1, v_2, \ldots, v_n] = g(v_1, g(v_2, \ldots, g(v_n, b)))
  \]
\end{theorem}
\begin{proof}
  We use the free monoid over constructors $a_1, a_2, \ldots$.  Furthermore,
  given a type $A$ with arbitrary elements $v_i$ we define the relation $R_A$ by
  \[
    m \mathrel{R_A} v \Longleftrightarrow m = a_i \land v = v_i \ \mbox{for some $i$}
  \]
  Since the type involves another quantified type $\beta$, we need to define
  a second relation $R_B$ where
  \[
    m \mathrel{R_B} d \Longleftrightarrow m = a_{i_1}\cdots a_{i_k} \land
    d = g(v_{i_1}, g(v_{i_2}, \ldots, g(v_{i_k}, b)))
  \]
  With these relations and the definition on of the logical predicate we get
  the following two properties.
  \begin{enumerate}
  \item $\forall m_1, m_2, v, d.\,  m_1 \mathrel{R_A} v \land m_2 \mathrel{R_B} d
    \Longrightarrow m_1\cdot m_2 \mathrel{R_B} g(v, d)$
  \item $\epsilon \mathrel{R_B} g$
  \end{enumerate}
  Since
  \[
    a_1 \cdots a_n \Vdash^{\alpha \mapsto R_A} [v_1, \ldots, v_n] \in [\mi{llist}\; \alpha]
  \]
  we can use the second and iterate the first property to conclude that
  \[
    a_1 \cdots a_n \mathrel{R_B} w \qquad \mbox{for $f\; g\; b\; [v_1, \ldots, v_n] \evalsto w$}
  \]
  By definition of $R_B$, this yields
  \[
    f\; g\; b\; [v_1, \ldots, v_n] = g(v_1, \ldots g(v_n, b))
  \]
  in the sense that both sides evaluate to $w$.  Because functions and values
  were chosen arbitrarily, this expresses the desired extensional equality.
\end{proof}

\section{Free Theorems Regarding Trees}

Consider
\begin{tabbing}
  \quad $\mi{lxrtree}\; \alpha = {\oplus}\{\ums{leaf} : \one, \ums{cons} : \mi{lxrtree}\; \alpha \fuse \alpha \fuse \mi{lxrtree}\; \alpha\}$ \\
  \quad $\mi{xlrtree}\; \alpha = {\oplus}\{\ums{leaf} : \one, \ums{cons} : (\mi{xlrtree}\; \alpha \twist \alpha) \fuse \mi{xlrtree}\; \alpha\}$ \\
  \quad $\mi{lrxtree}\; \alpha = {\oplus}\{\ums{leaf} : \one, \ums{cons} : \mi{lrxtree}\; \alpha \fuse (\alpha \twist \mi{xlrtree}\; \alpha)\}$
\end{tabbing}

Here are a few free theorems regarding such trees.  Further variations exist.

\begin{theorem}[]\mbox{}
  \label{thm:tree-traverse}
  \begin{enumerate}
  \item If $f : \forall \alpha.\, \mi{lxrtree}\; \alpha \rarrow \mi{llist}\; \alpha$
    then $f\; t$ lists the elements of $t$ following an \emph{inorder traversal}. 
  \item If $f : \forall \alpha.\, \mi{xlrtree}\; \alpha \rarrow \mi{llist}\; \alpha$
    then $f\; t$ lists the elements of $t$ following a \emph{preorder traversal}. 
  \item If $f : \forall \alpha.\, \mi{lrxtree}\; \alpha \rarrow \mi{llist}\; \alpha$
    then $f\; t$ lists the elements of $t$ following a \emph{postorder traversal}. 
  \end{enumerate}
\end{theorem}
\begin{proof}
  Trees, like lists, are purely positive types.  As such, we can prove an
  analogue of \autoref{lm:ordlist}.  We only show one of them, writing $t$ for
  tree values.
  \[
    \begin{array}{lcl}
      m \Vdash^S t \in [\mi{lxrtree}\; \alpha]
      & \Longleftrightarrow
      & m = \epsilon \land t = \ums{leaf}(\,) \\
      & & \null \lor \exists m_1, k, m_2.\, m = m_1 \cdot k \cdot m_2 \land v = \ums{node}(t_1, v, t_2) \\
      & & \quad \null \land m_1 \Vdash^S t_1 \in [\mi{lxrtree}\; \alpha]
          \land k \mathrel{S(\alpha)} v \land m_2 \Vdash^S t_2 \in [\mi{lxrtree}\; \alpha]
    \end{array}
  \]
\end{proof}



\section{From Ordered to Linear Types}

Exploring parametricity for \emph{linear} types instead of ordered ones is now a
rather straightforward change.  We conflate the left and right implication into
a single implication, and similarly for conjunction.
\[
  \begin{array}{lll|l}
    \mbox{ordered} & \mbox{linear} & \mbox{structural} & \mbox{values} \\\hline
    B \lover A & & \\
    & A \multimap B & A \arrow B & \lambda x.\, e\\
    A \lunder B & & \\\hline
    A \fuse B & & \\
    & A \tensor B & A \times B & (v_1, v_2) \\
    A \twist B & & \\\hline
    \one & \one & \one & (\,) \\
    {\oplus}\{\ell : A_\ell\} & {\oplus}\{\ell : A_\ell\} & {\oplus}\{\ell : A_\ell\}
    & \ell(v)
  \end{array}
\]
We see that in the transition from the linear to the structural case, no further
connectives collapse.  That's because we would still distinguish eager pairs
($A \times B$) from lazy records that we have elided from our development since
they do not introduce any fundamentally new ideas.

From the point of view of typing, the easiest change is to just permit the
silent rule of exchange
\[
  \infer[\ms{exchange}]
  {\Delta \mid \Gamma \semi \Omega_L\, (x:A)\, (y:B)\, \Omega_R \vdash e : C}
  {\Delta \mid \Gamma \semi \Omega_L\, (y:B)\, (x:A)\, \Omega_R \vdash e : C}
\]
The more typical change is to replace context concatenation
$\Omega_L\, \Omega_R$ with context merge $\Omega_L \bowtie \Omega_R$ which
allows arbitrary interleavings of the hypotheses.

Our definition of the logical predicates remains that same, except that we
assume that the algebraic structure parameterizing our definitions is a
\emph{commutative monoid}.  This immediately validates the rules of exchange and
the fundamental theorem goes through as before.

The results of exploiting the fundamental theorem to obtain parametricity results
are no longer as sharp.  For example:

\begin{theorem}
  If $\cdot \vdash e : \forall \alpha.\, \alpha \lolli \alpha \lolli \alpha \tensor \alpha$
  then $f$ is extensionally equal to $\lambda x.\, \lambda y.\, (x, y)$
  or $\lambda x.\, \lambda y.\, (y, x)$.
\end{theorem}
\begin{proof}
  By the fundamental theorem, we have
  \[
    \epsilon \Vdash e \in \llb \forall \alpha.\, \alpha \lolli \alpha \lolli \alpha \tensor \alpha \rrb
  \]
  Therefore $e \evalsto f$ and
  \[
    \epsilon \Vdash f \in [\forall \alpha.\, \alpha \lolli \alpha \lolli \alpha \tensor \alpha]
  \]
  We use a free commutative monoid with two generators, $a$ and $b$, arbitrary
  values $v$ and $w$ such that $a \mathrel{R} v$ and $b \mathrel{R} w$.  By the
  fundamental theorem:
  \[
    \epsilon \Vdash^{\alpha \mapsto R} f \in [\alpha \lolli \alpha \lolli \alpha \tensor \alpha]
  \]
  Applying this function to $v$ and $w$, we obtain that $f\, v\, w \evalsto p$
  and
  \[
    a \cdot b \Vdash^{\alpha \mapsto R} p \in [\alpha \tensor \alpha]
  \]
  This is true, again by definition, if for some $m$ and $k$ and $p_1$ and $p_2$ we have
  \[
    m \cdot k = a \cdot b \land p = (p_1,p_2) \land m \Vdash^{\alpha \mapsto R} p_1 \in [\alpha]
    \land k \Vdash^{\alpha \mapsto R} p_2 \in [\alpha]
  \]
  Further applying definitions, we get that for some $m$, $k$, $p_1$, and $p_2$, we have
  \[
    m \cdot k = a \cdot b \land m \mathrel{R} p_1 \land k \mathrel{R} p_2
  \]
  There are 4 ways that $a \cdot b$ could be decomposed into $m \cdot k$, but
  the definition of $R$ leaves only two possibilities: $m = a$, $k = b$, $p_1 = v$
  and $p_2 = w$ or $m = b$, $k = a$, $p_1 = w$ and $p_2 = v$.  Summarizing:
  either
  \[
    e\, v\, w \evalsto (v, w)
  \]
  or
  \[
    e\, v\, w \evalsto (w, v)
  \]
  which expresses that $e$ is extensionally equal to
  $\lambda x.\, \lambda y.\, (x, y)$ or $\lambda x.\, \lambda y.\, (y, x)$.
\end{proof}



\begin{theorem}
  If
  $\cdot \vdash e : \forall \alpha. \mi{list}\, \alpha \lolli \mi{list}\,
  \alpha$ then $e$ is extensionally equal to a permutation of the list elements.
\end{theorem}
\begin{proof}
  As in the proof of the related ordered theorem, we apply the fundamental
  theorem and then the definition for arbitrary values $v_i$ with
  $a_i \mathrel{R} v_i$ where $\alpha \mapsto R$, and the commutative monoid is
  freely generated from $a_1, a_2, \ldots$.

  Taking analogous steps to the ordered case, we conclude that
  $a_1 \cdots a_n = m_1 \cdots m_n$ modulo commutative (and associativity, as
  always) where each $m_i$ is a unique $a_j$.
\end{proof}

In the unrestricted case where various algebraic elements are fixed to be
$\epsilon$, we can only obtain that every element of the output list must be a
member of the input list, because those elements are in
$\epsilon \mathrel{R} v_i$.  We do not write out the details of this
straightforward adaptation of foregoing proofs.


\section{Related Work}


The most directly related work is Zhao et al.'s~\cite{zhao2010relational} open
logical relation for parametricity for a dual intuitionistic-linear polymorphic
lambda calculus.  In this work, they define an {\em open} logical relation that
includes an analog of typing contexts in the semantic model.  While our
development follows a similar structure, our resource algebraic account allows
us to eliminate spurious typechecking premises in definitions and permits a more
flexible range of substructural type systems.

Ahmed, Fluet, and Morrisett~\cite{ahmed2005step} introduce a logical relation
for substructural state via step-indexing, followed by ~\cite{ahmed2007l3} a
{\em linear language with locations} (L3) defined by a Kripke-style logical
relation to account for a language with mutable storage. However, the underlying
languages in these developments do not support parametric polymorphism. Ahmed,
Dreyer, and Rossberg later provide a logical relations account of a System
F-based language supporting imperative state update, and they demonstrate
representation independence results for this system~\cite{ahmed2009state}.  The
languages modeled in this body of work represent a specific point in the design
space with respect to imperative state update and references, as opposed to our
more general schema for substructural types in a functional setting. However,
Kripke-style logical relations that model a store as a partial commutative
monoid have some parallels to our development, and drawing out a more precise
relationship between these systems represents an interesting path of future
work.

Finally, there are a few developments that start from different settings but
develop semantics with similar properties. P\'{e}rez et al. develop logical
relations for linear session types~\cite{Perez12esop,Perez14ic} to establish
normalization results, but there is no account of parametricity.  The Iris
system for program reasoning via higher-order separation logic incorporates a
semantic model initially based on monoids~\cite{jung2015iris}, which is later
extended to more general resource algebras~\cite{jung2018iris}. Their
parameterization over resource algebras seems to work similarly to ours, but
towards the goal of program verification rather than type-based reasoning.  The
use of ``resource semantics'' more generally to account for the semantics of
substructural logics extends at least to Kamide~\cite{kamide2002kripke} and the
logic of bunched implications~\cite{o1999logic}, and similar ideas have recently
gained traction in the context of graded modal type systems~\cite{Vollmer25csl}.



\section{Conclusion}

We have provided an account of substructural parametricity including ordered,
linear, and unrestricted disciplines.  The fewer structural properties are
supported, the more precise the characterization of a function's behavior from
its type.  We have also implemented an ordered type checker using a
bidirectional type system with so-called additive contexts~\cite{Atkey18lics},
but the details are beyond the scope of this paper. Suffice it to say that all
the functions such as append, reverse, tree traversals, and fold can actually be
implemented in a variety of ways and are therefore not vacuous theorems.

The most immediate item of future work is to support general inductive and
coinductive types instead of purely positive recursive types.  This would allow
a new class of applications, including (productive) stream processing and
object-oriented program patterns.  We also envision an adjoint
combination of different substructural type systems~\cite{Jang24fscd},
extended to include exchange among the explicit structural rules.


\bibliography{fp,lr,lfs}

\begin{thebibliography}{10}

\bibitem{Abel16jfp}
Andreas Abel and Brigitte Pientka.
\newblock Well-founded recursion with copatterns and sized types.
\newblock {\em Journal of Functional Programming}, 26:e2, 2016.

\bibitem{ahmed2009state}
Amal Ahmed, Derek Dreyer, and Andreas Rossberg.
\newblock State-dependent representation independence.
\newblock {\em SIGPLAN Not.}, 44(1):340–353, January 2009.
\newblock \href {https://doi.org/10.1145/1594834.1480925}
  {\path{doi:10.1145/1594834.1480925}}.

\bibitem{ahmed2005step}
Amal Ahmed, Matthew Fluet, and Greg Morrisett.
\newblock A step-indexed model of substructural state.
\newblock In {\em Proceedings of the tenth ACM SIGPLAN international conference
  on Functional programming}, pages 78--91, 2005.

\bibitem{ahmed2007l3}
Amal Ahmed, Matthew Fluet, and Greg Morrisett.
\newblock L\^{} 3: a linear language with locations.
\newblock {\em Fundamenta Informaticae}, 77(4):397--449, 2007.

\bibitem{Atkey18lics}
Robert Atkey.
\newblock Syntax and semantics of quantitative type theory.
\newblock In Anuj Dawar and Erich Gr{\"a}del, editors, {\em 33rd Conference on
  Logic in Computer Science (LICS 2018)}, pages 56--65, Oxford, UK, July 2018.
  ACM.

\bibitem{Benton94csl}
P.~N. Benton.
\newblock A mixed linear and non-linear logic: Proofs, terms and models.
\newblock In Leszek Pacholski and Jerzy Tiuryn, editors, {\em Selected Papers
  from the 8th International Workshop on Computer Science Logic (CSL'94)},
  pages 121--135, Kazimierz, Poland, September 1994. Springer LNCS 933.
\newblock An extended version appears as Technical Report UCAM-CL-TR-352,
  University of Cambridge.

\bibitem{Caires10concur}
Lu{\'\i}s Caires and Frank Pfenning.
\newblock Session types as intuitionistic linear propositions.
\newblock In {\em Proceedings of the 21st International Conference on
  Concurrency Theory (CONCUR 2010)}, pages 222--236, Paris, France, August
  2010. Springer LNCS 6269.

\bibitem{Fu20lics}
Peng Fu, Kohei Kishida, and Peter Selinger.
\newblock Linear dependent type theory for quantum programming languages.
\newblock In {\em 34th Symposium on Logic in Computer Science (LICS 2020)},
  pages 440--453, Saarbr{\"u}cken, Germany, July 2020. ACM.

\bibitem{Gay10jfp}
Simon~J. Gay and Vasco~T. Vasconcelos.
\newblock Linear type theory for asynchronous session types.
\newblock {\em Journal of Functional Programming}, 20(1):19--50, January 2010.

\bibitem{Georges25popl}
A{\"\i}na~Linn Georges, Benjamin Peters, Laila Albeheiry, Leo White, Stephan
  Dolan, Richard~A. Eisenberg, Chris Casinghino, Fran{\c{c}}ois Pottier, and
  Derek Dreyer.
\newblock Data race freedom {\`a} la mode.
\newblock In {\em Principles of Programming Languages (POPL 2025)}, volume~9 of
  {\em Proceedings on Programming Languages}, pages 656--686. ACM, January
  2025.

\bibitem{Girard87tapsoft}
Jean-Yves Girard and Yves Lafont.
\newblock Linear logic and lazy computation.
\newblock In H.~Ehrig, R.~Kowalski, G.~Levi, and U.~Montanari, editors, {\em
  Proceedings of the International Joint Conference on Theory and Practice of
  Software Development}, volume~2, pages 52--66, Pisa, Italy, March 1987.
  Springer-Verlag LNCS 250.

\bibitem{Jang24fscd}
Junyoung Jang, Sophia Roshal, Frank Pfenning, and Brigitte Pientka.
\newblock Adjoint natural deduction.
\newblock In Jakob Rehof, editor, {\em 9th International Conference on Formal
  Structures for Computation and Deduction (FSCD 2024)}, pages 15:1--15:23,
  Tallinn, Estonia, July 2024. LIPIcs 299.
\newblock Extended version available as \url{https://arxiv.org/abs/2402.01428}.

\bibitem{jung2018iris}
Ralf Jung, Robbert Krebbers, Jacques-Henri Jourdan, Ale{\v{s}} Bizjak, Lars
  Birkedal, and Derek Dreyer.
\newblock Iris from the ground up: A modular foundation for higher-order
  concurrent separation logic.
\newblock {\em Journal of Functional Programming}, 28:e20, 2018.

\bibitem{jung2015iris}
Ralf Jung, David Swasey, Filip Sieczkowski, Kasper Svendsen, Aaron Turon, Lars
  Birkedal, and Derek Dreyer.
\newblock Iris: Monoids and invariants as an orthogonal basis for concurrent
  reasoning.
\newblock {\em ACM SIGPLAN Notices}, 50(1):637--650, 2015.

\bibitem{Kahn87}
Gilles Kahn.
\newblock Natural semantics.
\newblock In {\em Proceedings of the Symposium on Theoretical Aspects of
  Computer Science}, pages 22--39. Springer-Verlag LNCS 247, 1987.

\bibitem{kamide2002kripke}
Norihiro Kamide.
\newblock Kripke semantics for modal substructural logics.
\newblock {\em Journal of Logic, Language and Information}, 11:453--470, 2002.

\bibitem{Kanovich18ijcar}
Max Kanovich, Stepan Kuznetsov, Vivek Nigam, and Andre Scedrov.
\newblock A logical framework with commutative and non-commutative
  subexponentials.
\newblock In {\em International Joint Conference on Automated Reasoning (IJCAR
  2018)}, pages 228--245. Springer LNAI 10900, 2018.

\bibitem{Lambek58}
Joachim Lambek.
\newblock The mathematics of sentence structure.
\newblock {\em The American Mathematical Monthly}, 65(3):154--170, 1958.

\bibitem{Lattuada23oopsla}
Andrea Lattuada, Travis Hance, Chanhee Cho, Matthias Brun, Isitha Subasinghe,
  Yi~Zhou, Jon Howell, Bryan Parno, and Chris Hawblitzel.
\newblock {Verus}: Verifying {Rust} programs using linear ghost types.
\newblock In {\em Proceedings of the ACM on Programming Languages}, volume 7
  (OOPSLA 2023), pages 286--315, April 2023.
\newblock Extended version available as \url{https://arxiv.org/abs/2303.05491}.

\bibitem{Lorenzen23icfp}
Anton Lorenzen, Daan Leijen, and Wouter Swierstra.
\newblock {FP$^2$}: Fully in-place functional programming.
\newblock In {\em International Conference on Functional Programming (ICFP
  2023)}, Proceedings on Programming Languages, pages 275--304. ACM, August
  2023.

\bibitem{Lorenzen24pldi}
Anton Lorenzen, Daan Leijen, Wouter Swierstra, and Sam Lindley.
\newblock The functional essense of imperative binary search trees.
\newblock In {\em Programming Language Design and Implementation (PLDI 2024)},
  volume~8 of {\em Proceedings on Programming Languages}, pages 518--542. ACM,
  January 2024.

\bibitem{MacCaull98jsl}
Wendy MacCaull.
\newblock Relational semantics and a relational proof system for full {L}ambek
  calculus.
\newblock {\em Journal of Symbolic Logic}, 63(2):623--637, 1998.

\bibitem{mitchell1986representation}
John~C Mitchell.
\newblock Representation independence and data abstraction.
\newblock In {\em Proceedings of the 13th ACM SIGACT-SIGPLAN symposium on
  Principles of programming languages}, pages 263--276, 1986.

\bibitem{Nanevski08tocl}
Aleksandar Nanevski, Frank Pfenning, and Brigitte Pientka.
\newblock Contextual modal type theory.
\newblock {\em Transactions on Computational Logic}, 9(3), 2008.

\bibitem{o1999logic}
Peter~W O'Hearn and David~J Pym.
\newblock The logic of bunched implications.
\newblock {\em Bulletin of Symbolic Logic}, 5(2):215--244, 1999.

\bibitem{Perez12esop}
Jorge~A. P{\'e}rez, Lu{\'\i}s Caires, Frank Pfenning, and Bernardo Toninho.
\newblock Termination in session-based concurrency via linear logical
  relations.
\newblock In H.~Seidl, editor, {\em 22nd European Symposium on Programming},
  ESOP'12, pages 539--558, Tallinn, Estonia, March 2012. Springer LNCS 7211.

\bibitem{Perez14ic}
Jorge~A. P\'erez, Lu\'{\i}s Caires, Frank Pfenning, and Bernardo Toninho.
\newblock Linear logical relations and observational equivalences for
  session-based concurrency.
\newblock {\em Information and Computation}, 239:254--302, 2014.

\bibitem{Pfenning23coordination}
Frank Pfenning and Klaas Pruiksma.
\newblock Relating message passing and shared memory, proof-theoretically.
\newblock In S.~Jongmans and A.~Lopes, editors, {\em 25th International
  Conference on Coordination Models and Languages (COORDINATION 2023)}, pages
  3--27, Lisbon, Portugal, June 2023. Springer LNCS 13908.
\newblock Notes to an invited talk.

\bibitem{Polakow01phd}
Jeff Polakow.
\newblock {\em Ordered Linear Logic and Applications}.
\newblock PhD thesis, Department of Computer Science, Carnegie Mellon
  University, August 2001.

\bibitem{Polakow99tlca}
Jeff Polakow and Frank Pfenning.
\newblock Natural deduction for intuitionistic non-commutative linear logic.
\newblock In J.-Y. Girard, editor, {\em Proceedings of the 4th International
  Conference on Typed Lambda Calculi and Applications (TLCA'99)}, pages
  295--309, L'Aquila, Italy, April 1999. Springer-Verlag LNCS 1581.

\bibitem{Pruiksma18aun}
Klaas Pruiksma, William Chargin, Frank Pfenning, and Jason Reed.
\newblock Adjoint logic and its concurrent operational interpretation.
\newblock Unpublished manuscript, January 2018.

\bibitem{Reed07hylo}
Jason Reed.
\newblock Hybridizing a logical framework.
\newblock In {\em Proceedings of the International Workshop on Hybrid Logic
  (HyLo'06)}, pages 135--148. Electronic Notes in Theoretical Computer Science,
  v.174(6), 2007.

\bibitem{Reed10un}
Jason Reed and Frank Pfenning.
\newblock A constructive approach to the resource semantics of substructural
  logics.
\newblock Unpublished Manuscript, May 2010.
\newblock URL: \url{https://www.cs.cmu.edu/~jcreed/papers/rp-substruct.pdf}.

\bibitem{Reed09phd}
Jason~C. Reed.
\newblock {\em A Hybrid Logical Framework}.
\newblock PhD thesis, Carnegie Mellon University, September 2009.
\newblock Available as Technical Report CMU-CS-09-155.

\bibitem{Reynolds83ip}
John~C. Reynolds.
\newblock Types, abstraction, and parametric polymorphism.
\newblock In R.E.A. Mason, editor, {\em Information Processing 83}, pages
  513--523. Elsevier, September 1983.

\bibitem{Strachey67}
Christopher Strachey.
\newblock Fundamental concepts in programming languages.
\newblock {\em Higher-Order and Symbolic Computation}, 13:11--49, 2000.
\newblock Notes for lecture course given at the International Summer School in
  Computer Programming at Copenhagen, Denmark, August 1967.

\bibitem{Vollmer25csl}
Victoria Vollmer, Danielle Marshall, and Harley Eades, III.
\newblock A mixed linear and graded logic: Proofs, terms, and models.
\newblock In {\em 33rd Conference on Computer Science Logic (CSL 2025)}, pages
  32:1--32:21, Amsterdam, The Netherlands, February 2025. LIPIcs 326.

\bibitem{Wadler89fpca}
Philip Wadler.
\newblock Theorems for free!
\newblock In J.~Stoy, editor, {\em Proceedings of the 4th International
  Conference on Functional Programming Languages and Computer Architecture
  (FPCA'89)}, pages 347--359, London, UK, September 1989. ACM.

\bibitem{Zhao10aplas}
Jianzhou Zhao, Qi~Zhang, and Steve Zdancewic.
\newblock Relational parametricity for a polymorphic linear lambda calculus.
\newblock In K.~Ueda, editor, {\em 8th Asian Symposium on Programming Languages
  and Systems (APLAS 2010)}, pages 344--359. Springer LNCS 6461, 2010.

\bibitem{zhao2010relational}
Jianzhou Zhao, Qi~Zhang, and Steve Zdancewic.
\newblock Relational parametricity for a polymorphic linear lambda calculus.
\newblock In {\em Programming Languages and Systems: 8th Asian Symposium, APLAS
  2010, Shanghai, China, November 28-December 1, 2010. Proceedings 8}, pages
  344--359. Springer, 2010.

\end{thebibliography}

\end{document}